\documentclass[12pt]{article}  %% LaTeX2e
\usepackage{eucal,epsfig}      %% LaTeX2e
\def\be{\begin{equation}}
\def\ee{\end{equation}}
\def\l{\label}
\def\Im{\mbox{Im}}
\def\Re{\mbox{Re}}

\newcommand{\plot}[1]
{\begin{center} \epsfxsize=7cm
\vspace{-5mm}
\parbox{\epsfxsize}{\epsffile{#1}}
\vspace{5mm}
\end{center}}

\begin{document}
\begin{titlepage}
\title{\large\bf
ISOELECTRONIUM CORRELATIONS AS A NONLINEAR
TWO-DIMENSIONAL TWO-PARTICLE TUNNEL EFFECT}
%\title{\bf Isoelectronium correlations as a nonlinear
%two-dimensional two-particle tunnel effect}

\author{\bf A.K. Aringazin$^{1,3}$ and M.B. Semenov$^{2,3}$}

\date{\normalsize
{$^1$Department of Theoretical Physics,
Karaganda State University, Karaganda 470074 Kazakstan}\\
{\tt ascar@ibr.kargu.krg.kz}\\
{$^2$Penza State University,
40 Krasnaya St, Penza 440017 Russia}\\
{\tt physics@diamond.stup.ac.ru}\\
{$^3$Institute for Basic Research, P.O. Box 1577, Palm Harbor,}\\
{FL 34682, USA}\\
{\tt http://www.i-b-r.org}\\[0.5cm]
{August 1, 2000} }

\maketitle

\abstract{ In this paper, we study a nonlinear two-dimensional
two-particle quantum tunnel effect with dissipation in diatomic
H-H system. We use an instanton technique based on the path
integral approach, and present analytical solutions, in the
approximation of linearized reaction coordinates and harmonic
potential. It appears that the tunnel effect leads to a specific
correlation between two electrons interacting with each other by
Coulomb potential. This effect can be viewed as a mechanism
supporting the isoelectronium state recently proposed by Santilli
and Shillady within the framework of isochemical model of the
hydrogen molecule. We show that, under a condition of effective
two-dimensional motion, contribution of the quantum tunnel
correlated configurations is essential in the electron dynamics.
Temperature dependence of the probability of the two-electron
tunnel transitions per unit time has been studied.}
\end{titlepage}

\section{Introduction}

In a recent paper, Santilli and Shillady \cite{1} suggested a new
isochemical model of the hydrogen molecule characterized by a bond
at short distances of the two valence electrons into a singlet
quasi-particle state called isoelectronium. Namely, the
isochemical model suggests introducing of attractive short-range
Hulten potential interaction between the electrons, in addition to
the usual Coulomb repulsive potential. An origin of the
short-range attractive potential has been assumed due to a deep
overlapping of wavepackets of the electrons giving rise to
nonlinear and other effects.

Both the cases of unstable and stable isoelectronium corresponding
to the general four-body and the restricted three-body problems,
respectively, have been analyzed, in Born-Oppenheimer
approximation. Standard Boys-Reeves numerical three-body and
four-body calculations \cite{1}, exact analytic three-body
solution \cite{2}, and analytic four-body Ritz variational study
\cite{3} of the model have been carried out. The Hulten potential
$V_h$, which contains two parameters, appeared to lead to rather
complicated calculations so that the simplified potentials, the
Gaussian-screened-Coulomb potential $V_g$ and the
exponential-screened-Coulomb potential $V_e$, have been used in
the above papers, to mimic the Hulten potential at long and short
distance asymptotics. A comparison of the obtained results with
experimental data on H$_2$ molecule showed that the approximation
of stable pointlike isoelectronium (the three-body problem) does
not fit the data while for the unstable isoelectronium case (the
general four-body problem) exact representation of the
experimental binding energy and bond length of H$_2$ molecule can
be achieved by fitting the correlation length parameter of the
potential. The ground state energy based estimation \cite{3}
showed that the contribution of the attractive interelectron
potential $V_e$ (i.e., the contribution of the potential related
to the isoelectronium) should be about 1\%...6\% of that of the
usual repulsive interelectron Coulomb potential, to meet the
experimental data. In ref. \cite{3} certain relation between the
two parameters of the Hulten potential was established, thus only
one parameter has been used for the fitting. This relation arose
from consideration of the two-electron system governed by the
Hulten potential, with the assumption that the electron pair could
form one-level bound state with zero ground state energy.

The effect of electron correlations made by the introduced
attractive short-range interelectron potentials (i.e., the
unstable isoelectronium), with the correlation length parameter
value of about 0.01 bohr appeared to be helpful in achieving exact
representation of the experimental data on H$_2$ molecule. Also,
within the Boys-Reeves framework the isoelectronium based approach
appeared to provide an essentially (at least 1000 times) faster
computer calculations \cite{1}, in comparison to a standard C.I.
calculation. This approach has been extended to the case of other
diatomic molecules \cite{1} and water molecule \cite{4}.

In the present paper, we study two-electron correlations arising,
alternatively, from the {\it two-dimensional} tunnel effect with
dissipation which can take place in diatomic molecules subjected
to strong external electromagnetic fields. More specifically, we
take H$_2$ molecule as a simple example and assume that the
three-dimensional potential of the model can be effectively
reduced to a two-dimensional one under the action of external
fields. Also, we assume that the H$_2$ molecule is a part of
molecular association. Due to obtained results, the two-particle
tunnel effect leads to a contribution which resembles that of the
above isoelectronium correlations in two dimensions. However, we
should to emphasize that we do not introduce any attractive
interelectron potential. Namely, the characteristic two-electron
correlation arises naturally as a consequence of the
two-dimensional two-particle tunnel effect with dissipation.

Present consideration is strongly motivated by recent studies of a
light gas called magnegas$^{TM}$, consisting due to a standard
chemical analysis mainly of carbon monoxide CO (41\%) and hydrogen
H$_2$ (48\%), produced by Santilli's PlasmaArcFlow$^{TM}$ chemical
reactor \cite{5, 6}. Gas-chromatography mass-spectrometry and
infrared spectroscopy data obtained at room temperatures clearly
indicate presence of high molecular mass species (up to 1000
a.m.u.) in magnegas, in a macroscopic percentage. They have no
strong infrared signature of conventional chemical bonds, except
for that of carbon monoxide CO and carbon dioxide CO$_2$, which is
present in magnegas in small percentage. The GC-MS/IR search
results using a library of about 138,000 chemical species did not
indicate any matches with these high molecular mass species. The
main hypothesis on the origin of these species called {\it
magnecules} is that they are formed from the usual molecules (CO,
H$_2$, etc.) bonded to each other in some way, under the initial
action of strong external electromagnetic fields produced by
underwater arc in the PlasmaArcFlow reactor \cite{5}. We refer the
interested reader to refs. \cite{5,6} for more detailed
information on magnegas and PlasmaArcFlow reactor, and their
applications.

In the present paper, we conjecture that the two-dimensional
tunnel effects play an essential role at the stage of forming of
the above mentioned high molecular mass species (magnecules)
inside the reactor, where the influence of strong external
electromagnetic fields takes place. Physical and chemical
processes inside the reactor concern specific plasmachemical
studies and can be elaborated elsewhere. Here, we note only that
the importance of studying of the tunnel effects is grounded on
the fact that a tunnel effect is known to be as a mechanism in
forming of formaldehyde (CH$_2$=O) polymer structure.

We concentrate on the study of consequences implied by the
eventual 'magnetic freezing` (lowering of the effective
temperature due to the effect of external magnetic fields) of the
hydrogen molecule in molecular association. The magnetic freezing
is characterized by an effective reduction of the
three-dimensional potential to some two-dimensional one
(`polarizing of orbits'). We should to emphasize that the system
under study is not a usual H$_2$ molecule since we assume the
effective two-dimensional treatment of electron dynamics and the
presence of neighbor molecules (molecular association) serving as
a heat bath; herebelow, the system under study is referred to as
H-H system.

We use oscillator potential as an approximation to superposition
of Cou\-lomb potentials in the H-H system in the molecular
association. Such an approximation is naturally used in
low-temperature chemical kinetics, with parameters of the
oscillator potential being adjusted to reproduce a ground state
energy level. Also, we account for the interelectron interaction
by using the Coulomb potential. We note that the ground state
energy of the "magnetically frozen" H$_2$ molecule (i.e., the H-H
system) is lower than that of the ordinary H$_2$ molecule. As the
result the binding energy of the H-H system is bigger than that of
H$_2$ molecule. Detailed study of the effect of external magnetic
fields is out of the scope of the present paper and will be made
elsewhere.

Set up of the model is the following. We consider two-dimensional
(planar) motion of two electrons around fixed nuclei of two H
atoms. The electrons thus move in a nonlinear potential
$U(q_1,q_2)$, where the reaction coordinates $q_1$ and $q_2$ of
the electrons in the H-H system can be linearized in the first
order approximation as shown in Fig.~1.
%%Fig.~\ref{Lin1}.
The linearization assumes introducing of {\it four} centers,
instead of the original two centers (protons).

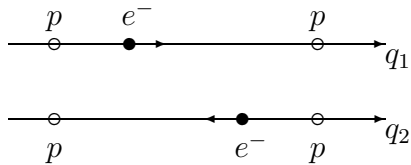
\begin{figure}[ht]              %Figure 1
\begin{center}
\unitlength=0.5mm
\begin{picture}(130,50)
\put(10,20){\vector(3,0){100}}  % horizontal axis lower
\put(70,20){\vector(-3,0){8}}   % horizontal vector
\put(20,10){$p$}                % horiz labels
\put(20,18){$\circ$}
\put(70,10){$e^-$}
\put(70,18){$\bullet$}
\put(90,10){$p$}
\put(90,18){$\circ$}
\put(110,15){$q_2$}
\put(10,40){\vector(3,0){100}}  % horizontal axis upper
\put(40,40){\vector(3,0){12}}   % horizontal vector
\put(20,45){$p$}                % horiz labels
\put(20,38){$\circ$}
\put(40,45){$e^-$}
\put(40,38){$\bullet$}
\put(90,45){$p$}
\put(90,38){$\circ$}
\put(110,35){$q_1$}
\end{picture}
\caption{The linearized reaction coordinates $q_1$ and $q_2$.}
\end{center}
\label{Lin1}
\end{figure}

The resulting approximate nonlinear potential $U(q_1,q_2)$
represents two-di\-men\-si\-onal potential surface modelling the
two-dimensional potential of the original system (two protons and
two electrons moving in the plane). This surface has the highest
points at a one-half of the proton-proton distance, and four
lowest points at the proton positions.

Dynamics of two interacting electrons in this two-dimensional
potential is due to the ordinary quantum mechanics to which the
energy barrier between the protons makes a specific contribution:
two-dimensional quantum tunnel effect of two-electron transitions.
The tunnel component of the probability of two-electron
transitions becomes essential in the case when the ordinary
overbarrier two-electron transitions are suppressed.

Clearly, the tunnel transitions can greatly affect both the
character and strength of the chemical bond in diatomic molecule.
As a consequence, they can give an essential contribution to
electron dynamics in the molecular association consisting of such
molecules.

We remark that usually a tunnel effect is considered for the case
of one particle in one dimension, which is a simplest case studied
to much extent. In this paper, we consider the tunnel effect with
{\it dissipation} for the case of {\it two interacting} particles
in a {\it two-dimensional} potential.

The tunnel correlation of electrons is considered on the
background of a formed two-dimensional potential of the H-H system
which is assumed to be a part of the molecular association.
Therefore we study temperature effects by introducing interaction
of the H-H system with a heat bath.

\section{The two-dimensional tunnel model}

Applications of the theory of quantum tunnelling with dissipation
\cite{7}-\cite{14} in chemical kinetics can be found in the
literature \cite{15}-\cite{17}. Below, we turn to determining of
the two-dimensional nonlinear potential for two interacting
electrons.

For the case of non-interacting electrons the potential energies of
first and second electrons, as functions of the reaction coordinates,
$q_1$ and $q_2$, are taken in the following form:
\begin{eqnarray}
\nonumber
U(q_1) = \frac{1}{2}\omega^2(q_1+a)^2\theta(-q_1)
       + \frac{1}{2}\omega^2(q_1-a)^2\theta(q_1),  \\
U(q_2) = \frac{1}{2}\omega^2(q_2-a)^2\theta(q_2)
       + \frac{1}{2}\omega^2(q_1+a)^2\theta(-q_2),
\end{eqnarray}
where $2a$ is the distance between two nuclei in the H-H system
(e.g., $2a\simeq 1.40$ bohr, for H$_2$ molecule in vacuum), mass
of electron is $m=1$, and $\omega$ is a frequency.
%(numerically, $\omega$ can be taken approximately as a frequency
%of rotation of electron in first Bohr orbit of H atom).
In Fig.~2,
%%Fig.~\ref{Lin2},
the potentials $U(q_1)$ and $U(q_2)$ are shown.

\begin{figure}[ht]            %Figure 2
\begin{center}
\unitlength=0.5mm
\begin{picture}(200,80)
%left figure
\put(40,20){\vector(0,0){60}} % vertical axis
\put(0,40){\vector(3,0){80}}  % horizontal axis
\put(25,55){\oval(30,30)[b]}  % left oval
\put(55,55){\oval(30,30)[b]}  % right oval
\put(15,35){\small $-a$}      % left horiz label
\put(55,35){\small $a$}       % right horiz label
\put(80,37){\small $q_1$}     % right horiz label
\put(42,75){\small $U(q_1)$}   % vert label
\put(32,48){\vector(3,0){15}}  % horizontal vector
\put(32,46){$\bullet$}         % horizontal vector
%right figure
\put(140,20){\vector(0,0){60}} % vertical axis
\put(100,40){\vector(3,0){80}} % horizontal axis
\put(125,55){\oval(30,30)[b]}  % left oval
\put(155,55){\oval(30,30)[b]}  % right oval
\put(115,35){\small $-a$}      % left horiz label
\put(155,35){\small $a$}       % right horiz label
\put(180,37){\small $q_2$}     % right horiz label
\put(142,75){\small $U(q_2)$}     % vert label
\put(147,48){\vector(-3,0){15}}   % horizontal vector
\put(145,46){$\bullet$}           % horizontal vector
\end{picture}
\caption{The potentials $U(q_1)$ and $U(q_2)$ as functions
of linearized coordinates.}
\end{center}
\label{Lin2}
\end{figure}
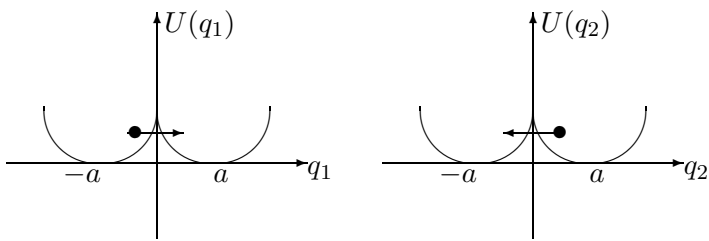

The interaction between two electrons is taken in the following
form:
\be\l{2}
V_{int}(q_1,q_2) = - \frac{1}{2}\alpha (q_1-q_2)^2,
\ee
where $\alpha$ is a positive parameter, and $(q_1-q_2)$ is
distance between the electrons. Note that $V_{int}$ has the form
of {\it attractive} harmonic potential. Indeed, in the $xy$-plane
the $x$-component of the mutual distance of two interacting
electrons in two dimensions are much bigger
than the $y$-component, $q_0\gg a$, as shown in Fig.~3 %Fig.~\ref{Tran3}

\begin{figure}[ht]                   %Figure 3
\begin{center}
\unitlength=0.7mm
\begin{picture}(110,70)
\put(60,10){\vector(0,0){60}}        % vertical axis
\put(0,40){\vector(3,0){120}}        % horizontal axis
\put(2,34){\small $-\frac{q_0}{2}$}  % left horiz label
\put(5,48){$\bullet$}                % left horiz label
\put(110,43){\small $\frac{q_0}{2}$} % right horiz label
\put(110,30){$\bullet$}              % right horiz label
\put(120,35){\small $q_x$}           % right horiz label
\put(62,70){\small $q_y$}            % vert label
\put(62,50){\small $q_{2y}$}         % vert label
\put(50,30){\small $q_{1y}$}         % vert label
\end{picture}
\caption{Transition of two interacting electrons.}
\end{center}
\label{Tran3}
\end{figure}
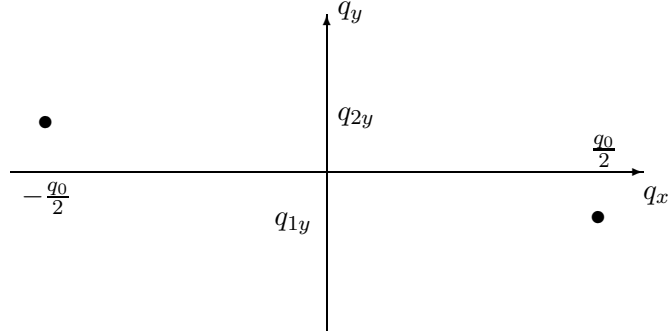

Thus, the interaction potential can be expanded in small parameter
$(q_{1y}-q_{2y})^2/q_0^2$, where $q_{1y}$ and $q_{2y}$ are
coordinates of tunneling. The Coulomb repulsion potential between
the electrons, with electric charges $-e$, is then
\begin{eqnarray}
\nonumber
V_{C} = \frac{e^2}{\varepsilon_0|q|} =
\frac{e^2}{\varepsilon_0\sqrt{q_0^2+(q_{1y}-q_{2y})^2}} \\
\simeq \frac{e^2}{\varepsilon_0q_0}
-\frac{e^2}{2\varepsilon_0q_0}\frac{(q_{1y}-q_{2y})^2}{q_0^2},
\l{3}
\end{eqnarray}
where $\varepsilon_0$ is the dielectric constant. The negative
(second) term in the series expansion (\ref{3}) is the effective
attractive potential (\ref{2}) so that the constant $\alpha$ is
found as
\be\l{4}
\alpha = \frac{e^2}{2\varepsilon_0q_0}.
\ee
The first term, $e^2/(\varepsilon_0q_0)$, is responsible for
repulsion between the electrons, and is constant along the
reaction coordinates so that it can be absorbed by redefining
$U(q_1)$ and $U(q_2)$. Obviously, the interaction potential $V_C$
is always repulsive but one can see that only its effective {\it
attractive} part given by Eq. (\ref{2}), as a function of the
reaction coordinates, appears to contribute to the reaction
dynamics.

The following remark is in order. An {\it attractive} character of
the interelectron potential $V_{int}(q_1,q_2)$ resembles the
attractive interelectron Hulten potential $V_h(r)$ of the
isochemical model \cite{1}. The difference is that $V_{int}$ is
formulated in terms of reaction coordinates, has a harmonic form,
and naturally arises as an effective long-range potential in the
two-dimensional approach. So, $V_{int}$ can be viewed as a
potential essentially supporting the isoelectronium state in two
dimensions.

Thus, the general form of the total potential is a sum of $U(q_1)$,
$U(q_1)$, and $V_{int}(q_1,q_2)$,
$$
\tilde U(q_1,q_2)= U(q_1)+U(q_2)+V_{int}(q_1,q_2),
$$
which we rewrite as
\be\l{5}
U(q_1,q_2) = \frac{2\tilde U(q_1,q_2)}{\omega^2}
= (q_1+a)^2\theta(-q_1)\theta(q_1)
+ (q_2-a)^2\theta(q_2)\theta(-q_2)
-\frac{\alpha^*}{2}(q_1-q_2)^2.
\ee
Here, we have denoted
$$
\alpha^* = \frac{2\alpha}{\omega^2},
$$
which is a dimensionless parameter, $\alpha^* <1$. In Fig.~4,
%%Fig.~\ref{Lev4}
a section of the potential (\ref{5}), at some energy level, is
schematically depicted.
In Figs.~5    %%\ref{3D5}
     and 6,   %%\ref{3D6}
three-dimensional plots of the potential, at two different values
of the renormalized interelectron coupling parameter, $\alpha^*$ =
0.1 (weak coupling) and $\alpha^*$ = 0.5 (strong coupling), are
shown. Here, the ovals corresponding to sections of the potential
by a horizontal plane at some energy level are depicted for the
reader convenience. We recall that $q_1$ and $q_2$ are the
reaction coordinates (not coordinates of real configuration space
of the system).

\begin{figure}[ht]                   %Figure 4
\begin{center}
\unitlength=1mm
\begin{picture}(110,80)
\put(60,0){\vector(0,0){80}}         % vertical axis
\put(0,40){\vector(3,0){120}}        % horizontal axis
\put(2,35){-$\frac{a}{1-\alpha^*}$}  % left horiz label
\put(35,41){-$a$}                    % left horiz label
\put(38,23){\oval(8,8)}              % left small oval
\put(38,23){4}                       % left small oval number
\put(5,70){\oval(14,14)}             % left big oval
\put(5,70){3}                        % left big oval number
\put(105,45){$\frac{a}{1-\alpha^*}$}  % right horiz label
\put(80,37){$a$}                      % right horiz label
\put(80,60){\oval(8,8)}               % right small oval
\put(80,60){2}                        % right small oval number
\put(110,10){\oval(14,14)}            % right big oval
\put(110,10){1}                       % right big oval number
\put(120,35){$q_1$}                   % right horiz label
\put(62,80){$q_2$}                   % vert label
\put(62,70){$\frac{a}{1-\alpha^*}$}  % vert label
\put(57,60){$a$}                     % vert label
\put(60,20){-$a$}                    % vert label
\put(50,10){-$\frac{a}{1-\alpha^*}$} % vert label
\end{picture}
\caption{Level lines (1,2,3,4) of the potential $U(q_1,q_2)$;
$\alpha^*$ is the renormalized coupling parameter.}
\end{center}
\label{Lev4}
\end{figure}
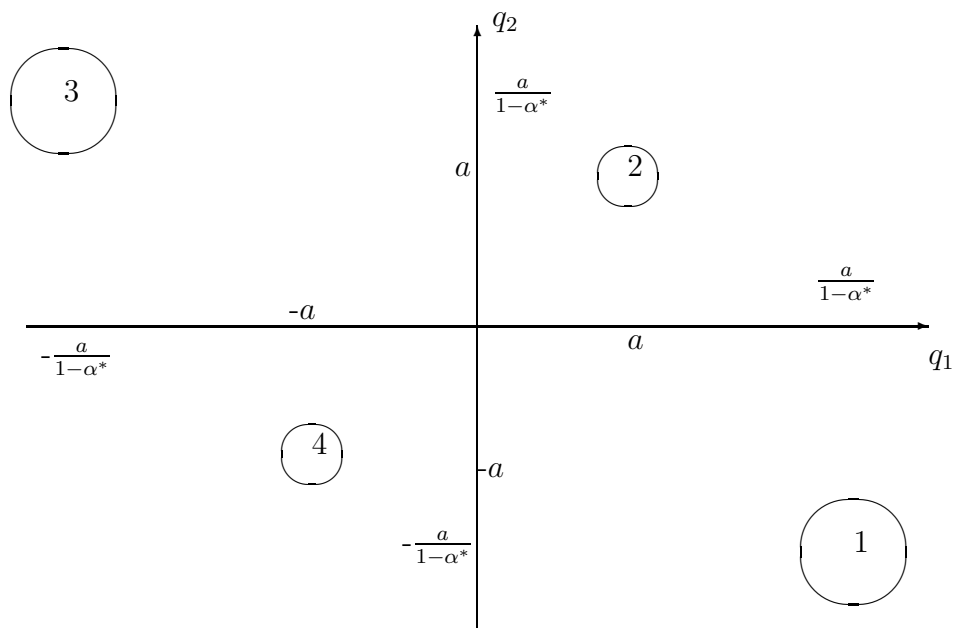

\begin{figure}[ht] %Figure 5 (eps-figure ibr10_5.eps)
\plot{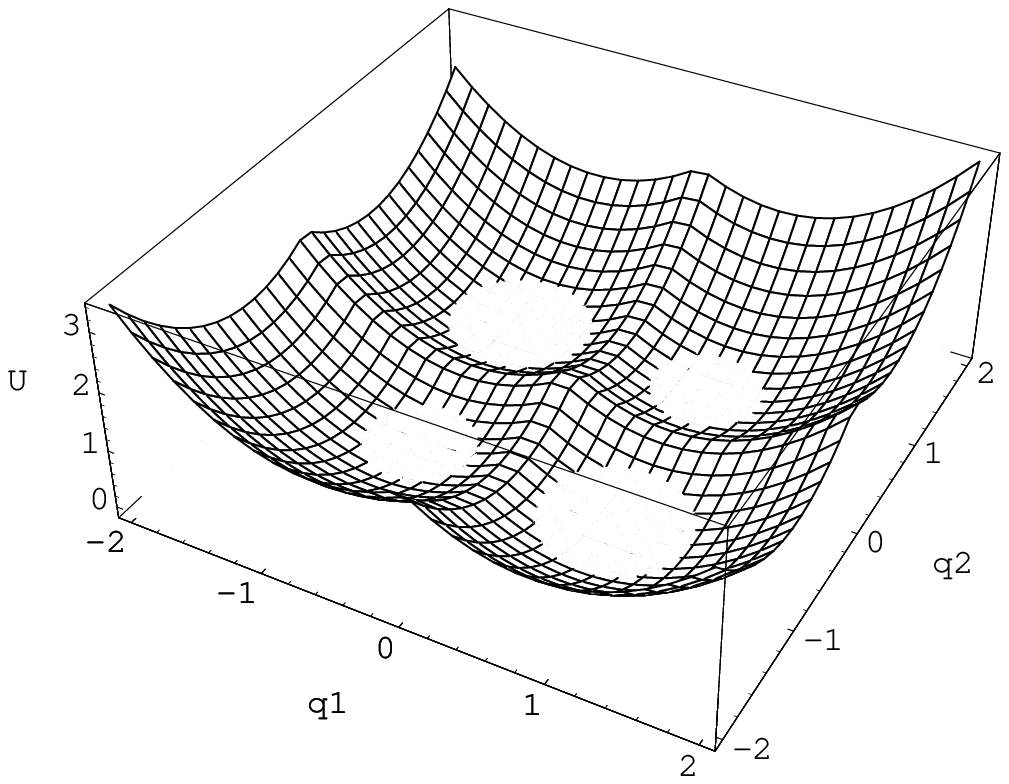}
\caption{A three-dimensional plot of the potential $U(q_1,q_2)$;
$\alpha^*$ = 0.1, $a$ = 0.7.}
\label{3D5}
\end{figure}

\begin{figure}[ht] %Figure 6 (eps-figure ibr10_6.eps)
\plot{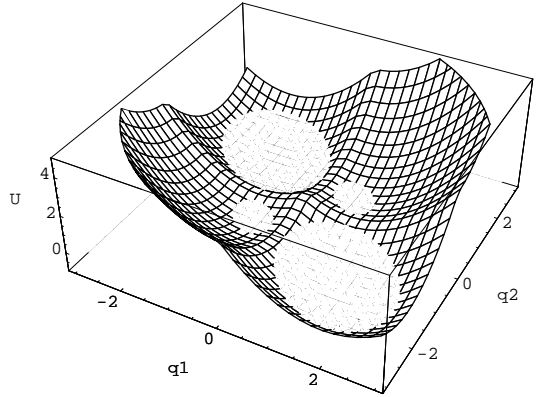}
\caption{A three-dimensional plot of the potential $U(q_1,q_2)$;
$\alpha^*$ = 0.5, $a$ = 0.7.}
\label{3D6}
\end{figure}

\section{Probability of the two-electron transition}

Since the hydrogen molecule is assumed to be a part of an
association of molecules (magnecule), we introduce interaction of
the H-H system with external oscillators, a heat bath. Dynamics of
the heat bath is defined by the oscillator Hamiltonian
\be\l{6}
H_{hb} = \frac{1}{2}\sum\limits_{i}(p_i+\omega_i^2Q_i^2).
\ee
We assume that each of the electrons linearly interacts
with the oscillators, namely, the interaction potentials are
\be\l{7}
V^{(1)}_{e-hb}(q_i,Q_i) = q_1\sum\limits_{i}C_iQ_i,
\qquad
V^{(2)}_{e-hb}(q_i,Q_i) = q_2\sum\limits_{i}C_iQ_i,
\ee
The probability of the two-electron transition per unit time
is given by
\be\l{8}
\Gamma = 2T \frac{\Im Z}{\Re Z},
\ee
where, as usual, the most important part is the exponential part
of $\Gamma$.
At zero temperature, for the case of metastable levels we have
\be\l{9}
\Gamma = -2\, \Im E, \qquad  E=E_0 - \frac{i}{2}\Gamma,
\ee
from which Eq. (\ref{8}) follows as the generalization
to the case of finite temperature $T$. Indeed,
\be\l{10}
\Gamma = \frac{2\sum\limits_i \exp (-E_{0i}/T) \Im E_i}
              { \sum\limits_i \exp (-E_{0i}/T)}
=        \frac{2T \Im \sum\limits_i \exp (-E_{i}/T)}
              {\Re \sum\limits_i \exp (-E_{i}/T)}
= 2T \frac{\Im Z}{\Re Z}.
\ee
Here, $i$ runs over energy levels of the metastable states, and
$Z$ is the statistical sum of the total system. The imaginary part
of $Z$ corresponds to a decay of the energy levels.

To calculate $\Gamma$ it is convenient to represent $Z$ as
a path integral \cite{7}-\cite{14}
\be\l{11}
Z = \prod\limits_i \int Dq_1Dq_2DQ_i \exp [-S(q_1,q_2,Q_i)],
\ee
where $S$ is the action of the total system. An imaginary part of
$Z$ corresponds to the decay of the initial energy levels. One can
perform functional integral over $Q_i$ exactly, and obtain
$$
Z = \prod\limits_i \int Dq_1Dq_2 \exp [-S(q_1,q_2)],
$$
where
\be\l{12}
S(q_1, q_2) = \int\limits_{-\beta/2}^{\beta/2} d\tau
\Bigl\{
\frac{1}{2}\dot{q}_1^2 + \frac{1}{2}\dot{q}_2^2 +V(q_1, q_2))
+
\ee
$$
\frac{1}{2}\int\limits_{-\beta/2}^{\beta/2} d\tau'
D(\tau-\tau')(q_1(\tau) +q_2(\tau))(q_1(\tau') +q_2(\tau'))
\Bigr\},
$$
where
\be\l{13}
D(\tau) = \frac{1}{\beta}\sum\limits_{n=-\infty}^{\infty}
D(\nu_n)e^{i\nu_n\tau}
\ee
is Green function for the oscillators, with
\be\l{14}
D(\nu_n) = - \sum\limits_i \frac{C_i^2}{\omega_i^2+\nu_n^2},
\ee
$\beta = \hbar/k_BT$ is the inverse temperature parameter,
and $\nu_n$ is Matsubara's frequency.

The quasiclassical trajectory minimizing the action $S$
in two-dimensional space is defined by the equations of motion,
\be\l{15}
-\ddot{q}_1 +\Omega_0^2q_1 + \alpha_1q_2
+ \int\limits_{-\beta/2}^{\beta/2} d\tau'
K(\tau-\tau')(q_1(\tau')+q_2(\tau')) + \omega^2a\theta(-q_1)
- \omega^2a\theta(q_1) = 0
\ee
and
\be\l{16}
-\ddot{q}_2 +\Omega_0^2q_2 + \alpha_1q_1
+ \int\limits_{-\beta/2}^{\beta/2} d\tau'
K(\tau-\tau')(q_1(\tau')+q_2(\tau')) - \omega^2a\theta(q_2)
+ \omega^2a\theta(-q_2) = 0.
\ee
Here, the kernel $K$ is defined by
\be\l{17}
K(\tau) = \frac{1}{\beta}\sum\limits_{n=-\infty}^{\infty}
\xi_n e^{i\nu_n\tau},
\ee
where $\xi_n$ is determined from the redefined Eq. (\ref{14}),
namely,
\be\l{18}
D(\nu_n) = - \sum\limits_i \frac{C_i^2}{\omega_i^2}+\xi_n.
\ee
This redefinition has been made in order to extract a zero mode term.

We seek for solutions of the set of Eqs. (\ref{15}) and (\ref{16})
in the form of Fourier series expansion in frequencies $\nu_n$,
\be\l{19}
q_1 = \frac{1}{\beta}\sum\limits_{n=-\infty}^{\infty}
q_n^{(1)}e^{i\nu_n\tau},
\qquad
q_2 = \frac{1}{\beta}\sum\limits_{n=-\infty}^{\infty}
q_n^{(2)}e^{i\nu_n\tau}.
\ee
The renormalized frequency, $\Omega_0$ and renormalized coupling
constant $\alpha_1$ are defined by
\be\l{20}
\Omega_0^2 = \omega^2 - \sum\limits_i\frac{C_i^2}{\omega_i^2} -\alpha,
\qquad
\alpha_1 = \alpha - \sum\limits_i\frac{C_i^2}{\omega_i^2}.
\ee

Inserting Eqs. (\ref{19}) into the set of equations
(\ref{15}) and (\ref{16}), we get the following equations:
\begin{eqnarray}
\nonumber
q_0^{(1)} + q_0^{(2)} = \frac{4\omega^2a\varepsilon}{\Omega_0^2+\alpha_1},
\\
q_0^{(1)} - q_0^{(2)} = -\frac{2\omega^2a\beta}{\Omega_0^2-\alpha_1}
+\frac{8\omega^2\tau_0}{\Omega_0^2-\alpha_1},
\l{21}
\end{eqnarray}
for the case $n=0$, and
\begin{eqnarray}
\nonumber
q_n^{(1)} + q_n^{(2)} =
\frac{4\omega^2a(\sin\nu_n\tau_1 -\sin\nu_n\tau_2)}
{\nu_n(\nu_n^2+\Omega_0^2+\alpha_1+2\xi_n)},
\\
q_n^{(1)} - q_n^{(2)} =
\frac{4\omega^2a(\sin\nu_n\tau_1 +\sin\nu_n\tau_2)}
{\nu_n(\nu_n^2+\Omega_0^2-\alpha_1)},
\l{22}
\end{eqnarray}
for the case $n\not=0$.
Here, we have denoted
\be\l{23}
\varepsilon = \tau_1 - \tau_2, \qquad
\tau_0 = \frac{1}{2}(\tau_1+\tau_2),
\ee
and $\tau_1$ and $\tau_2$ are the values of time $\tau$
at which first and second electron, respectively,
passes through the maximum of the potential barrier.
Namely, $\tau_1$ and $\tau_2$ are determined from the following equations:
\be\l{24}
q_1(\tau_1)=0, \qquad q_2(\tau_2)=0.
\ee
These two equations can be used to change arguments of the above
functions $\theta$. As the result, the $\theta$ functions become
depending on the time values $\tau_1$ and $\tau_2$, instead of the
coordinates $q_1$ and $q_2$, and the Eqs. (\ref{15}) and
(\ref{16}) take a {\it linear} form. Note that, quasiclassically,
the time values $\pm\tau_1$ and $\pm\tau_2$ correspond to the
times when first and second electron, respectively, are at the top
of the potential barrier.

Inserting the trajectories defined by Eqs. (\ref{19}), (\ref{21}),
and (\ref{22}) into Eqs. (\ref{12}), we obtain the quasiclassical
(instanton) action in the form
$$
S= \frac{8\omega^4a^3\tau_0}{\Omega_0^2-\alpha_1}
-\frac{4\omega^4a^2\varepsilon^2}{\beta(\Omega_0^2+\alpha_1)}
-\frac{16\omega^4a^2\tau_0^2}{\beta(\Omega_0^2-\alpha_1)}
$$
\be\l{25}
-\frac{32\omega^4a^2}{\beta}
\sum\limits_{n=1}^{\infty}
\left[
\frac{\sin^2\nu_n\tau_0 +\cos^2\nu_n\varepsilon/2}
{\nu_n^2(\nu_n^2+\Omega_0^2-\alpha_1)}
+
\frac{\sin^2\nu_n\varepsilon/2 +\cos^2\nu_n\tau_0}
{\nu_n^2(\nu_n^2+\Omega_0^2+\alpha_1+2\xi_n)}
\right].
\ee
The quasiclassical action $S$ describes probability of the
two-particle tunnel transition per unit time, with an exponential
accuracy. For the case of {\it antiparallel} motion of the
electrons, the above instanton action $S$, at $\xi_n=0$ (the heat
bath effects are ignored here), takes the form
$$
S = -2\omega a^2
\left\{
|\varepsilon|(1- \frac{1}{1-\alpha^*})
+\frac{\sinh(|\varepsilon|\sqrt{1-\alpha^*})}{(1-\alpha^*)^{3/2}}
-\sinh|\varepsilon|
\right.
$$
\be\l{26}
+\frac{\cosh(|\varepsilon|\sqrt{1-\alpha^*})+1}{(1-\alpha^*)^{3/2}}
\cdot
\frac{\cosh((\beta^*-\tau)\sqrt{1-\alpha^*})-\cosh(\beta^*\sqrt{1-\alpha^*})}
{\sinh(\beta^*\sqrt{1-\alpha^*})}
\ee
$$
\left.
+\frac{(\cosh\varepsilon - 1)(\cosh(\beta^*-\tau)+\cosh\beta^*)}
      {\sinh\beta^*}
\right\}.
$$
Here, we have denoted
$$
\alpha^* = \frac{2\alpha}{\omega^2},
\quad
\beta^* = \frac{\beta\omega}{2},
\quad
\varepsilon = (\tau_1-\tau_2)\omega,
\quad
\tau = (\tau_1+\tau_2)\omega,
$$
and the parameters $\varepsilon$ and $\tau$ obey the following
set of equations:
$$
-\sinh\varepsilon [\coth\beta^* + \cosh\tau\coth\beta^* - \sinh\tau]
+\frac{1}{1-\alpha^*}\sinh(\varepsilon\sqrt{1-\alpha^*})
[\coth(\beta^*\sqrt{1-\alpha^*})
$$
$$
-\cosh(\tau\sqrt{1-\alpha^*})
\coth(\beta^*\sqrt{1-\alpha^*})+\sinh(\tau\sqrt{1-\alpha^*})] =0,
$$
\be\l{27}
-1 - \frac{1}{1-\alpha^*}
+ (\cosh\varepsilon-1)(\sinh\tau\coth\beta^*-\cosh\tau)
+\cosh\varepsilon
\ee
$$
+\frac{1}{1-\alpha^*}\left\{
[\cosh(\varepsilon\sqrt{1-\alpha^*})+1]
[\sinh(\tau\sqrt{1-\alpha^*})\coth(\beta^*\sqrt{1-\alpha^*})
\right.
$$
$$
\left.
-\cosh(\varepsilon\sqrt{1-\alpha^*})]
-\cosh(\varepsilon\sqrt{1-\alpha^*})
\right\}.
$$
Solution of the Eq. (\ref{27}) is
\be\l{28}
\varepsilon = (\tau_1-\tau_2)\omega = 0,
\quad
\alpha<\frac{1}{2}\omega^2,
\quad
\forall \beta,
\ee
$$
\tau_1=\tau_2 = \frac{\tau}{2\omega} = \frac{\beta}{4}.
$$
At low temperatures, $\omega\beta \gg 1$, we then have
\be\l{29}
e^{-\tau\sqrt{1-\alpha^*}} \simeq \alpha^*(1-\alpha^*)^\gamma
\left\{
1- (1-\alpha^*)^\gamma(\frac{\alpha^*}{1-\sqrt{1-\alpha^*}} -1)
\right\}^{-1},
\ee
$$
e^\varepsilon \simeq \frac{1}{1-\alpha^*}
\frac{(\alpha^* - e^{-\tau\sqrt{1-\alpha^*}})}{e^{-\tau\sqrt{1-\alpha^*}}},
$$
where
$$
\gamma = \frac{\sqrt{1-\alpha^*}}{1-\sqrt{1-\alpha^*}}.
$$
The approximate solution (\ref{29}) is valid at
$$
\frac{1}{4}<\frac{2\alpha}{\omega^2}<1
$$
and $\beta > \beta_c$, where the critical temperature is
\be\l{30}
\beta_c =
- \frac{1}{\omega\sqrt{1-\alpha^*}}\ln
\left\{
\frac{\alpha^* (1-\alpha^*)^\gamma}
{1-(1-\alpha^*)^\gamma(\frac{\alpha^*}{1-\sqrt{1-\alpha^*}}-1)}
\right\}.
\ee
As one can see, there arises a characteristic critical
temperature, $\beta_c= \hbar/(k_BT_c)$, of the system, and it
depends mainly on the value of the interelectron coupling
parameter $\alpha^*$.

At $\varepsilon =0$,  for the case of
symmetric two-dimensional potential the solution (\ref{28}) yields
\be\l{31}
S = \frac{4\omega a^2}{(1-\alpha^*)^{3/2}}
\tanh \frac{\omega\sqrt{1-\alpha^*}}{4}\beta
\ee
At $\varepsilon \not=0$, the resulting expression for the action $S$
is cumbersome so that we do not represent it here.
However, it can be shown that, at $\beta>\beta_c$,
\be\l{32}
S_{(\varepsilon=0)} < S_{(\varepsilon\not=0)}
\ee
so that the value $\varepsilon =0$ minimizes the action. As the
result, the values $\varepsilon \not=0$ correspond to {\sl
splitted} trajectories which appear to be unstable;
see Fig.~7. %%Fig.~\ref{Spl7}.

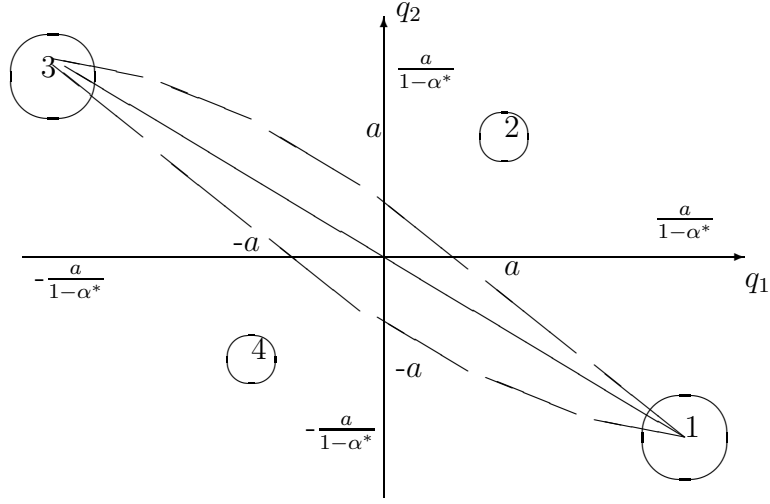
\begin{figure}[ht]                   %Figure 7
\begin{center}
\unitlength=0.8mm
\begin{picture}(110,80)
\put(60,0){\vector(0,0){80}}         % vertical axis
\put(0,40){\vector(3,0){120}}        % horizontal axis
\put(2,35){-$\frac{a}{1-\alpha^*}$}  % left horiz label
\put(35,41){-$a$}                    % left horiz label
\put(38,23){\oval(8,8)}              % left small oval
\put(38,23){4}                       % left small oval number
\put(5,70){\oval(14,14)}             % left big oval
\put(3,70){3}                        % left big oval number
\put(105,45){$\frac{a}{1-\alpha^*}$}  % right horiz label
\put(80,37){$a$}                      % right horiz label
\put(80,60){\oval(8,8)}               % right small oval
\put(80,60){2}                        % right small oval number
\put(110,10){\oval(14,14)}            % right big oval
\put(110,10){1}                       % right big oval number
\put(120,35){$q_1$}                   % right horiz label
\put(62,80){$q_2$}                   % vert label
\put(62,70){$\frac{a}{1-\alpha^*}$}  % vert label
\put(57,60){$a$}                     % vert label
\put(62,20){-$a$}                    % vert label
\put(47,10){-$\frac{a}{1-\alpha^*}$} % vert label
\put(110,10){\line(-5,3){103}}        %slope line
\put(110,10){\line(-5,4){15}}         %upper dash line
\put(92,24){\line(-5,4){15}}
\put(74,38){\line(-5,4){15}}
\put(56,52){\line(-5,3){15}}
\put(38,63){\line(-5,2){15}}
\put(20,70){\line(-5,1){15}}
\put(110,10){\line(-5,1){15}}         %lower dash line
\put(92,14){\line(-5,2){15}}
\put(74,21){\line(-5,3){15}}
\put(56,31){\line(-5,4){15}}
\put(38,45){\line(-5,4){15}}
\put(20,60){\line(-5,4){15}}
\end{picture}
\caption{A schematic picture of the tunnel trajectories
at $\varepsilon=0$ (solid line)
and at $\varepsilon \not=0$ (dash lines).}
\end{center}
\label{Spl7}
\end{figure}

The instanton solution for $\varepsilon=0$ corresponds to a
strongly correlated antiparallel motion of the two tunnelling
electrons. Namely, the two electrons {\it simultaneously} pass the
top points of the potentials. This correlation of the two
electrons depends on the value of the electron-electron
(attraction) coupling constant $\alpha^*$ naturally arising within
the framework of our model.

Such a correlation resembles the isoelectronium correlation
introduced in ref. \cite{1} where the correlation was introduced
within a different approach and is governed by the short-range
attractive electron-electron Hulten potential.

\section{Conclusions}

At $\beta>\beta_c$, the basic single trajectory is splitted
into two trajectories as shown in Fig.~7 %Fig.~\ref{Spl7}.
(dash lines). In contrast to the case of parallel transition, this
splitting (bifurcation) takes place at any values of the
parameters of the potential $U(q_1,q_2)$. Also, at
$\beta>\beta_c$, we have $S_{(\varepsilon\not=0)} >
S_{(\varepsilon=0)}$ so that the trajectory with $\varepsilon=0$
make a biggest contribution. At $\beta<\beta_c$, the two
trajectories become a single trajectory characterized by
$q_1=-q_2$.

We see that in contrast to the case of one tunnelling particle at
which only one trajectory (instanton) is realized, in the case of
two tunnelling particles in the two-dimensional potential the
situation is more complicated. As we have shown in general there
are two types of trajectories, a single basic trajectory and
splitted (degenerate) trajectory, both making contribution to the
tunnel effect. The splitted trajectory is characterized by a
nonsimultaneous ($\tau_1\not=\tau_2$) transition of the two
particles through the top points of the barriers, so that the
time-correlation between the particles' motion is lost. It is
highly remarkable that in the case of symmetric two-dimensional
potential such splitted trajectories in both the cases of parallel
and antiparallel motion of the two particles are {\sl unstable}.
At small $\alpha^*$ and $\beta<\beta_c$, the degenerate trajectory
is not realized so that there is only the basic single trajectory
($q_1=-q_2$). This trajectory is characterized by a simultaneous
($\tau_1=\tau_2$) transition of the two particles through the top
points of the barriers, so that we observe a specific correlation
between the electrons.

The chosen form of the interelectron interaction
does not affect the motion of the center of mass
($q_1=q_2$). So, for the parallel transition along the basic
trajectory ($q_1=q_2$) the quasiclassical (instanton) action
does not depend on the coupling parameter. In this case,
\be\l{33}
S = 4\omega a^2 \tanh\frac{\omega\beta}{4}.
\ee
Since the motion with maximal value of the relative coordinate,
i.e. $q_1=-q_2$, is energetically preferable it becomes clear why
for the parallel transition along the degenerate trajectory the
action decreases with the increase of the coupling parameter,
while for the antiparallel transition along the degenerate
trajectory the action increases with the increase of the coupling
parameter.

Thus, in the case of antiparallel tunnel transition of two
interacting electrons the preferable trajectory is a single basic
one which is characterized by strong (isoelectronium-like)
correlation between the electrons ($\tau_1=\tau_2$). Here, the
uncorrelated motion ("decay" of the isoelectronium-like state) is
{\sl suppressed} because it makes bigger contribution to the
instanton action.

As to a heat bath, it appears to be possible to study analytically
the effect of, for example, one local mode $\omega_L$ to the
two-electron transition probability. The result is that the heat
bath does not change the qualitative conclusions made above.
Temperature dependence of the action $S$ for the antiparallel
motion case ($\varepsilon=0$) at two different values of the
renormalized coupling parameter, $\alpha^*=0.1$ and
$\alpha^*=0.5$, is shown in
Fig.~8
%%Fig.~\ref{Tem8}.

\begin{figure}[ht]             %Figure 8
\begin{center}
\unitlength=0.5mm
\begin{picture}(110,80)
\put(10,10){\vector(0,0){60}}  % vertical axis
\put(10,10){\vector(1,0){100}} % horizontal axis
\put(13,65){$S/4\omega a^2$}   % vert labels
\put(3,24){1}
\put(3,44){2}
\put(115,5){$\beta\omega/4$}   % horiz labels
\put(10,3){0}
\put(20,3){1}
\put(30,3){2}
\put(40,3){3}
\put(50,3){4}
\put(60,3){5}
\put(70,3){6}
\put(80,3){7}
\put(15,15){$\circ$}      %plot(a)
\put(25,22){$\circ$}
\put(35,25){$\circ$}
\put(45,27){$\circ$}
\put(55,28){$\circ$}
\put(65,29){$\circ$}
\put(75,30){$\circ$}
\put(85,30){$\circ$}
\put(95,30){$(a)$}
\put(15,19){$\bullet$}    %plot(b)
\put(25,33){$\bullet$}
\put(35,40){$\bullet$}
\put(45,47){$\bullet$}
\put(55,52){$\bullet$}
\put(65,55){$\bullet$}
\put(75,56){$\bullet$}
\put(85,56){$\bullet$}
\put(95,56){$(b)$}
\end{picture}
\caption{The action ($\varepsilon=0$, antiparallel motion)
as a function of inverse temperature $\beta = \hbar/k_BT$;
(a) $\alpha^*$ = 0.1, (b) $\alpha^*$ = 0.5.}
\end{center}
\label{Tem8}
\end{figure}
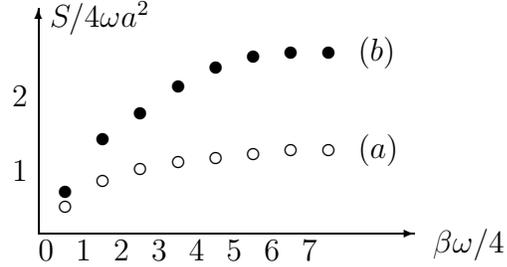

In this paper, we used the assumption that dynamics of the two
interacting electrons can be taken approximately as a
two-dimensional. This is due to the viewpoint \cite{1} that strong
external magnetic field confines, to some degree, the usual
three-dimensional motion of the electrons so that the
consideration of {\it two-dimensional} effects in molecules, as
well as in associations of molecules, becomes important. The main
two-dimensional effect, namely, the two-electron tunnel effect
with dissipation, for the antiparallel transition, has been
studied in this paper.

By using the approximations of one-instanton linearized reaction
coordinates and an ideal gas of instanton-antiinstanton pairs
($\beta\omega\ll 16U_{barrier}/\omega$), we have shown that the
{\it correlated isoelectronium-like motion} of the two tunnelling
electrons in two-dimensional H-H system with dissipation takes
place, and appears to be the {\it most stable} configuration.
Thus, the tunnel correlations can be viewed as a mechanism
supporting the idea of isoelectronium states introduced by
Santilli and Shillady \cite{1}.

Detailed study of the effect of strong external magnetic fields,
as well as the implications of the results of this paper, will be
made in a subsequent paper.

Application of the presented instanton formalism to carbon
monoxide molecule $CO$ is of much interest as well. Also, study of
the transition rates, $\Gamma = e^{-S}$, for the case of
association of molecules (intramolecular tunnelling) could be
important in investigating electron charge distribution and bonds
in the association of molecules (magnecules).

\section*{Acknowledgment}
The authors are grateful to Yu.I. Dakhnovskii for a support of this work.

\end{document}